# DRL-Based RAT Selection in a Hybrid Vehicular Communication Network

Badreddine Yacine YACHEUR
LaBRI CNRS UMR5800, Univ. Bordeaux
Bordeaux INP, F-33400
Talence, France
byacheur@u-bordeaux.fr
[0000-0002-3679-3319]

Toufik AHMED
LaBRI CNRS UMR5800, Univ. Bordeaux
Bordeaux INP, F-33400
Talence, France
tad@labri.fr
[0000-0002-9245-0759]

Mohamed MOSBAH
LaBRI CNRS UMR5800, Univ. Bordeaux
Bordeaux INP, F-33400
Talence, France
mosbah@labri.fr
[0000-0001-6031-4237]

***Abstract*—Cooperative intelligent transport systems rely on a set of Vehicle-to-Everything (V2X) applications to enhance road safety. Emerging new V2X applications like Advanced Driver Assistance Systems (ADASs) and Connected Autonomous Driving (CAD) applications depend on a significant amount of shared data and require high reliability, low end-to-end (E2E) latency, and high throughput. However, present V2X communication technologies such as ITS-G5 and C-V2X (Cellular V2X) cannot satisfy these requirements alone. In this paper, we propose an intelligent, scalable hybrid vehicular communication architecture that leverages the performance of multiple Radio Access Technologies (RATs) to meet the needs of these applications. Then, we propose a communication mode selection algorithm based on Deep Reinforcement Learning (DRL) to maximize the network's reliability while limiting resource consumption. Finally, we assess our work using the platooning scenario that requires high reliability. Numerical results reveal that the hybrid vehicular communication architecture has the potential to enhance the packet reception rate (PRR) by up to 30% compared to both the static RAT selection strategy and the multi-criteria decision-making (MCDM) selection algorithm. Additionally, it improves the efficiency of the redundant communication mode by 20% regarding resource consumption.**

***Keywords*—*Hybrid vehicular network, ITS-G5, C-V2X, RAT selection, Deep reinforcement learning***

## I. Introduction

Vehicle-to-everything applications are classified into three categories according to their requirements. Day 1 V2X applications are for anticipatory driving. Day 2 V2X applications aim to develop collaborative perception and awareness, and Day 3 V2X applications tend to reach fully automated accident-free driving. New generation technologies such as IEEE 802.11bd, 3GPP C-V2X, and 5G NR are being developed to meet the requirements of these last two categories. However, none of them could meet the required QoS [1]. In vehicular networks, the observed performance of a technology typically changes dynamically over time due to frequently changing channel states arising from high node mobility. Moreover, each technology has its weaknesses and strengths, even with ITS-G5 and LTE-V2X PC5 technologies with similar functionalities. For example, in [2], a study shows that LTE-V2X PC5 achieves higher reliability at considerable communication distances, and ITS-G5 is better at close communication distances. Studies from paper [3] show that the improvement gain of LTE-V2X PC5 over 802.11p is minor in the case of a short platoon (e.g., up to five tracks). Nevertheless, LTE-V2X PC5 is more performant with long platoons due to a better link budget. So, getting the required performance to lunch Day 2 and Day 3 applications with only one technology is challenging.

This paper proposes a hybrid vehicular network architecture that emphasizes using ITS-G5 and LTE-V2X technologies simultaneously. RAT selection is made with a DRL-based algorithm that considers the channel quality and application requirements to determine the optimal communication mode. Furthermore, we deal with a high mobility communication nature in VANETs where the broadcast success is difficult to estimate, and the link quality is variable. So, based only on local observations, the feedback of the Reinforcement Learning (RL) environment is hard to define. Thus, to get a better estimate of the feedback of each agent, we use the multi-channel communication aspect of ITS-G5 to get the reception acknowledgment [4]. To get seamless connectivity and transparency aspects of the architecture, we follow the same architecture pattern as [1] by adding a hybrid communication management layer to the ITS-station architecture. The remaining of this paper is structured as follows; in Section II, we debrief some related works on hybrid vehicular networks and the use of RL in decision problems. Section III discusses the proposed hybrid vehicular architecture and the DRL-based communication mode selection algorithm. Later in Section IV, we present the simulation setup and discuss the performance results. Finally, Section V concludes the paper and highlights future work.

## II. Related works

Works on hybrid vehicular networks are often addressed to enable applications that are part of the Day 3 V2X applications. Paper [1] introduced a protocol stack with a new layer, the Hybrid Communications Management (HCM) layer. The HCM determines a dynamic mapping between the application requirements and the communication technologies. They also investigated single-technology and multi-technology networks to increase the reliability or the network's throughput. However, no concrete evaluation was realized regarding multi-technology selection. In the paper [5], another hybrid vehicular network architecture approach was proposed. It combines Dedicated Short-Range Communication (DSRC) technology-enabled ad-hoc network and an infrastructure-based LTE network. They structured the protocol stack of this architecture as a generic framework called CellCar. This framework defines a dynamic RAT selection and a dynamic communication management mechanism. The evaluation of this architecture shows that a hybrid architecture can give higher reliability and lower delays. However, this work does not include the defined LTE-V2X PC5, nor does it consider the simultaneous multi-technology use as discussed in the previous work. Decision problems like RAT selection can be solved using static service-aware RAT selection, a deterministic decision-making method like Multi-Criteria Decision-Making algorithms (MCDM), or Reinforcement learning algorithms. In [6], the authors proposed a DIstributed and Context Aware Radio Access Technology selection (DICART) framework for vehicular networks using the Technique for Order Preference by Similarity to an Ideal Solution (TOPSIS) to choose the best one that ensures an Always Best Connected (ABC) facility. Paper [7] proposes a RL framework based on slow fading parameters and statistical information to perform V2V communication mode selection and power adaptation in 5G communication networks. Other works like [8] address network selection and resource allocation problems in IoT using RL and deep Multi-Agent RL (DMARL). To our knowledge, our work is the first to evaluate the efficiency of the simultaneous use of multi-RAT with an RL-based RAT selection algorithm in a hybrid vehicular communication network. Furthermore, we compare our DRL-based selection algorithm with baselines (single RAT) and other selection algorithms like static RAT selection algorithm [9] and TOPSIS-based MCDM algorithm [6].

## III. Hybrid Vehicular Communication Architecture and the DRL-Based RAT Selection

This section presents the hybrid vehicular communication architecture and its RAT selection strategy. We adopt a flat hybrid architecture [10] where each vehicle is equipped with ITS-G2 and C-V2X. Hybrid communication is done using three communication modes. The first is a single communication mode, where only one communication link is used to transmit the message, as is usually done in a legacy vehicular communication network. The second communication mode is a redundant mode where a message is duplicated to achieve higher reliability [1]. Where the third communication mode is the division mode. The message is sent through two RATs as two independent transmissions to enhance the transmission throughput [1]. Furthermore, our hybrid vehicular communication architecture is scalable to other RAT standards, as depicted in Fig. 1, such as IEEE 802.11bd, LTE-V2X UU, and 5G NR. Each technology has a dedicated Radio Resource Management (RRM) entity in the management layer to monitor information about the channel quality. The Access Layer monitors the communication interface and provides information on the network state to the RRM, such as frame transmission statistics, Signal to Noise plus Interference Ratio (SNIR), and other channel load indicators. The hybrid communication layer encompasses all control operations, including communication mode selection, communication technology configuration, and message processing before and after the transmission and reception of a packet. With this layer, hybrid communication is transparent to the facilities layer.

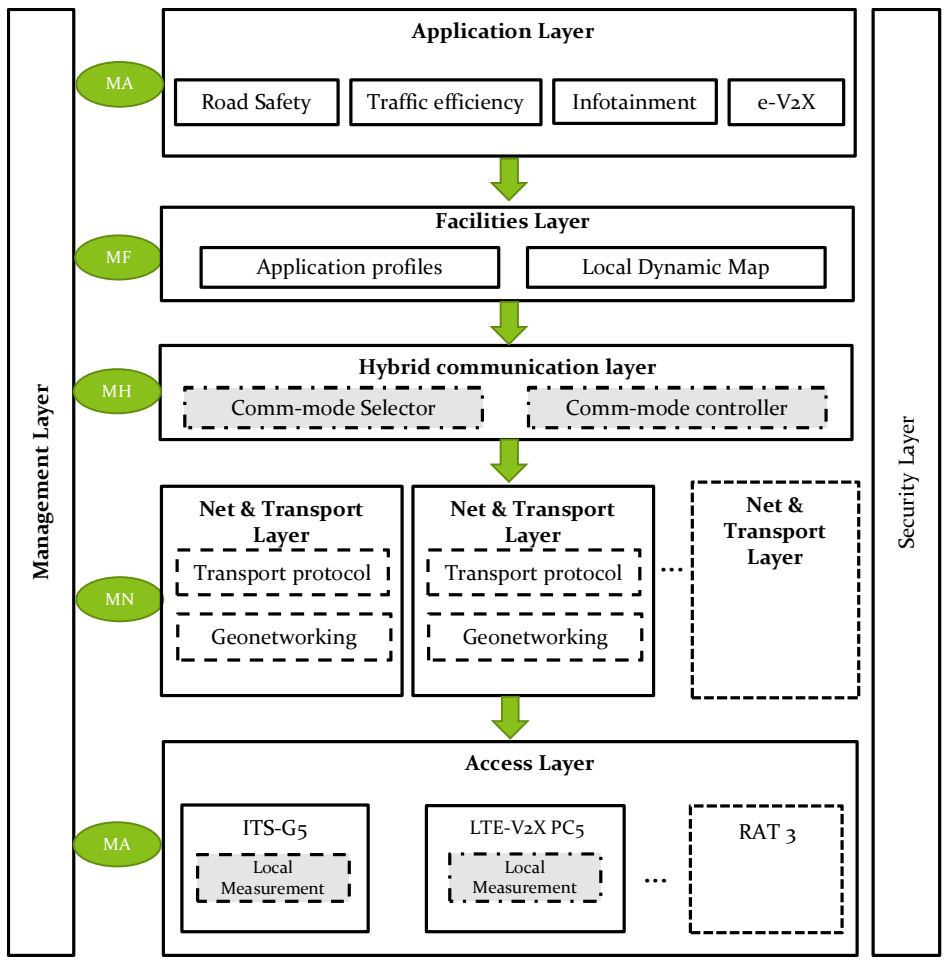

Fig. 1. The hybrid communication layer in the ITS-S protocol stack

We use Deep Reinforcement Learning (DRL) to determine the appropriate communication mode based on network quality and application needs. The objective of RL is to teach the vehicle how to deal with the stochastic nature of the vehicular communication network in order to fulfill the application's requirements. Markov Decision Process (MDP) is generally used as a formal means for RL. However, we cannot solve our problem using an MDP because the communication mode selection problem does not have a finite number of states. To this end, we are using DRL. In our DRL depiction, each vehicle is considered an agent. The environment state in the state space $\mathcal{S}$ is composed of six elements:

$$s_t = [SNIR_{ITS-G}\ , SNIR_{LTE}, PRR_{ITS-G}\ , PRR_{LTE}, L, R]$$

Where $SNIR_{ITS-G}$ and $SNIR_{LTE}$ are the measured SNIR of each RAT network. $PRR_{ITS-G5}$ and $PRR_{LTE}$ are the calculated packet reception ratio of each RAT. L and R represent the service's latency and reliability requirements, respectively. These parameters cannot be obtained before sending a message via the transmitter [1]. The only way to get the value of these parameters is to get them at the reception of a message. So, we consider three different environment states to have accurate state representations. $s_t$ as the actual environment state, $s'_t$ as the state of the environment after the vehicle receives the first message from another vehicle in the environment. And the next state is the state of the environment after the vehicle receives the latest message before it sends a new one $s_{t+1}$.

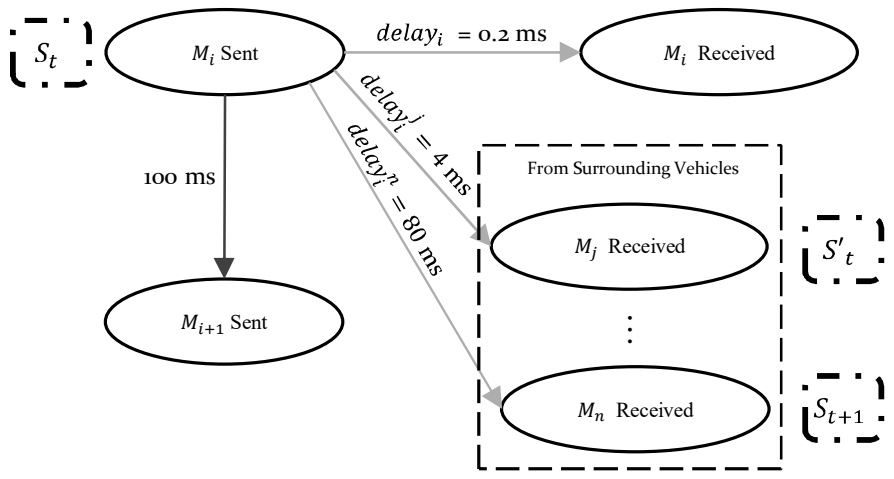

Fig. 2. Different state representations of the environment

Using these three representations, an agent chooses an action based on $s_t$. It receives an evaluation of its action based on the earliest link quality measurement, which is the one measured at $s'_t$. And it updates the current state with the values of $s_{t+1}$. The relation between each state is depicted in Fig. 2. Our discrete action space is given as follows:

$$A = \begin{cases} 0, & Single_{ITS-G5} \\ 1, & Single_{LTE} \\ 2, & Hybrid_{Redandant} \\ 3, & Hybrid_{Division} \end{cases}$$

The reward is composed of three parts as follows:

$$R_{V_i} = \begin{cases} \frac{1}{2}\sum_{j=0}^{n} ReceptionState[V_i][V_j], \ i \neq j \\ + \\ \frac{\alpha}{2} PS \\ + \\ \frac{\beta}{2} LQ \end{cases}$$

The first part concerns the message reception evaluation, which checks if all surrounding vehicles have received the message successfully. This kind of acknowledgment is delivered using the multi-channel aspect of ITS-G5 to avoid congesting the channel with message reception acknowledgments as used in [4]. Every 100 milliseconds, an agent sends a message. And knowing that the $delay_i$ (End-to-End latency), represented in Fig. 2, is always less than 100 milliseconds. Each agent can receive an acknowledgment from its surrounding vehicles before sending another message. These acknowledgments are represented as a vector named the reception evaluation vector, which is updated every time a message is received. The vector contains reception reports and a success reception (SR) counter. If a message is received twice by a vehicle, a value of "two" is reported, and if it was not received, a value of "zero" is reported. Otherwise, the message is considered perfectly received, and a value of "one" is reported. The SR counter represents the number of perfectly received messages, which is incremented by one if all reception reports in the vector are equal to "one." The second part of the reward function is about performance satisfaction (PS). It guides the agent in choosing the best communication mode to satisfy the application's requirements. The third part gives feedback to the agent about the link quality (LQ) enhancement. It compares the current state ($s_t$) SNIR values with the earlier next state ($s'_t$) SNIR values.

Double deep Q-learning algorithm combines Q-learning and deep learning. It employs a neural network known as the behavior network, which approximates the Q-value $Q(s_t, a_t, \omega)$ using its parameters ($\omega$). These parameters are constantly adjusted to match the optimal strategy $\pi^*$ by learning from uncorrelated experiences data. The algorithm also uses a second neural network with the same structure and initial parameters $\omega'$ as the approximation neural network (a.k.a., target network). During the learning process, explained in Algorithm 1, we randomly sample a minibatch of tuples and updates the neural network parameters according to a variant of the stochastic gradient descent (SGD) method, named mini-batch SGD. This mini-batch is picked from a replay buffer (aka., experience replay) that collects and stores $< s_t, a_t, r_t, s_{t+1} >$ tuple in every iteration. The mini-batch is used to update $\omega$, thus minimizing the following loss function:

$$Loss(\omega) = [r_t + \gamma \max_{a'} Q(s_t, a', \omega') - Q(s_t, a_t, \omega)]^2 / M \quad (1)$$

Where M is the size of the sampled mini-batch and $Q(s_t, a_t, \omega')$ is calculated using the Bellman optimality equation [11] (2) and the target network output.

$$Q_{\pi^*}(s_t, a_t, \omega) = E[r_t + \gamma max_{a'} Q_{\pi^*}(s'_t, a'_t, \omega) \mid s_t, a_t] \quad (2)$$

**Algorithm 1**:Double deep Q-learning Algorithm for communication mode selection in a hybrid vehicular communication network

```
Input
Initialized behavior network and target network parameters.
Initialized replay buffer of size L and a batch size of 64.
ϵ = 1, and ϵ_decrement = 10^-5
γ = 0.99
Output
Learned agents
Strat
for episode = 1 to 1000 do {
  // Initialize the environment and get initial state observations
  o_t = [SNIR_ITS-G5, SNIR_LTE, PRR_ITS-G5, PRR_LTE, L, R]
  while (reception_vector[SR] < target) do {
    //According to ϵ choose the action and decrement ϵ
    a_t = { random action
          { a_t = argmax_a_t Q(s_t, a, ω) }
    ϵ = ϵ − ϵ_decrement
    // Perform action a_t, and get the immediate s'_t, r_t using s'_t
    If (all reported values in reception_vector = 1)
      reception_vector [SR]++
    clean(reception_vector)
    // Get the observation of the next state s_t+1
    O_t = O_t+1
    store_transition (< s_t, a_t, r_t, s_t+1 >)
    // Train the model if the buffer size is > 64
    sample-random mini-batch of {s_t, a_t, r_t, s_t+1}
    // Calculate prediction using the mini-batch
    Q_target → { r_t
               { r_t + γ max_a' Q(s_t, a', ω') }
    Q_behavior → Q(s_t, a_t, ω)
    Perform mini-batch SGD using the loss function
    Update behavior network parameter
    Update target network parameters every T round
  } end while
  reception_vector[SR] = 0
} end for
```

## IV. Performance Evaluation

To assess our hybrid communication architecture, we simulate the ITS-G5 and LTE-V2X PC5 technologies using the OMNeT++ framework, SUMO traffic simulator, Artery Framework [12], and SimuLTE [13]. We compare our DRL-based selection approach with a simple legacy approach (i.e., a single RAT vehicular network, ITS-G5, or LTE-V2X), a static RAT selection strategy, and an MCDM mode selection strategy that leverages TOPSIS, as used in [6].

### A. *Simulation scenario*

The ITS station (ITS-S) architecture supports multiple RATs in the Access layer. However, hybrid vehicular communication is not supported by the OMNeT++ simulator. So, to use multiple RATs simultaneously, we implemented the hybrid communication layer using simple OMNeT++ modules, as depicted in Fig. 3. Next, to assess our architecture, we lunch a platooning service [14] within a five vehicles platoon. We consider a two-way highway scenario with two lanes in each direction and a length of two kilometers. Two base stations have been deployed as infrastructure to support the LTE-V2X PC5 (mode 3) standard. There are no structures in the vicinity of the highway. The spacing between vehicles in the platoon is 10 meters, and each vehicle follows the Cooperative Adaptive Cruise Control (CACC) car following model as established by the SUMO simulator. The maximum speed limit for platoon vehicles is 10 meters per second and 20 meters per second for all other vehicles. Each vehicle is equipped with ITS-G5 and LTE-V2X PC5, as described in Section III. All parameters are summarized in TABLE I.

TABLE I. STANDARDS CONFIGURATION

| ***Parameter*** | ***ITS-G5*** | ***LTE-V2X*** |
|---|---|---|
| Access mechanism | EDCA | SPS (mode 3) |
| Access channel | 5875 MHz – 5885 MHz | 5895MHz – 5905 MHz |
| Transmitter power | 23 dBm | 23 dBm |
| Receiver sensitivity | -85 dBm | -85 dBm |
| Energy detection | -85 dBm | -85 dBm |
| Background noise | -90dBm | -110dBm |
| Propagation model | Constant Speed Propagation | Jakes |

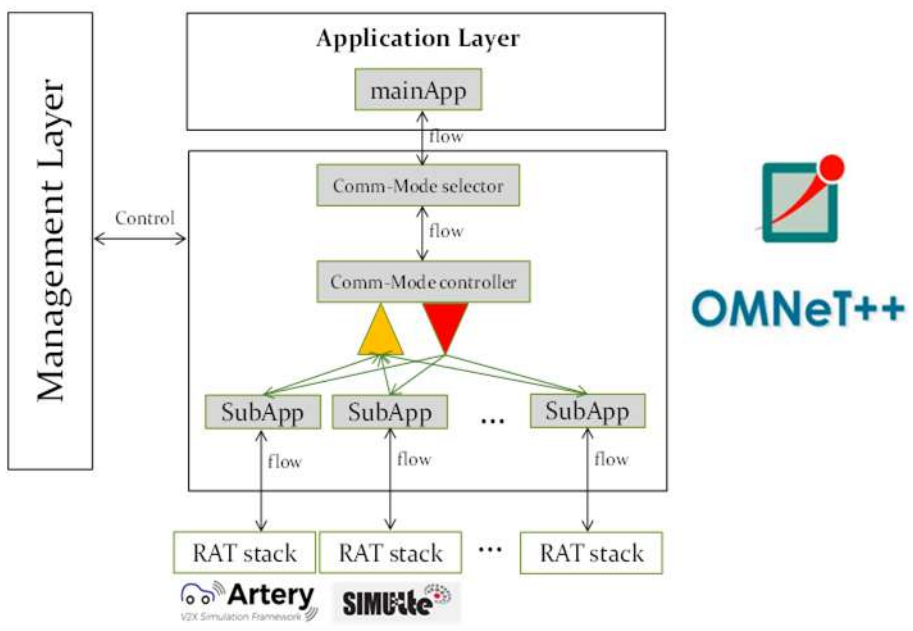

Fig. 3. Simulation environment

## B. *Double deep Q-learning algorithm parameters*

The OMNeT++ simulator is implemented using C++. So, to make the interaction between the DRL model and the simulation of the vehicular communications, we implemented the Double deep Q-learning algorithm using the libtorch library. Libtorch is the C++ interpretation of the well-known python machine learning framework Pytorch. The behavior and the target networks are composed of four layers: an input layer with six neurons, two hidden layers with 256 neurons, and an output layer with three neurons. The activation function in each layer is a ReLu function. The ADAM optimizer and the MSE loss function update the neural network parameters. The learning rate is set to 0.0005, $\gamma$ is set to 0.99, and the epsilon decay is set to $10^{-5}$. The maximum size of the replay buffer is set to $10^{6}$, and the mini-batch size is set to 64. In our experiments, a game ends when the SR counter reaches the target of 100 received messages, not to have long games and to give the agent the time to learn. In a game, one step is equivalent to sending a V2V message. The PRR is calculated using equation (3). Where the $N_{sent\ messages}$ is the number of sent messages to reach 100.

$$PRR_{game} = \frac{SR\ target}{N_{sent\ messages}} \tag{3}$$

## C. *Simulation results*

We employ three performance metrics to evaluate the performance of our hybrid vehicular communication network and to assess our work. The average PRR, the percentage of duplicate messages when redundant mode is enabled. Moreover, to evaluate the convergence performance of our proposed selection algorithm, we present the average expected reward per game to show the convergence of the algorithm. First, we present the evaluation of our DRL-based communication mode selection algorithm. Fig. 4 represents the agent behavior in a medium-congested network. Fig. 4 (a) shows that the reward converges at game number 600. This convergence is followed by an excellent reliability improvement, as depicted in Fig. 4 (b). We notice that in the first 400 games, the reliability of the transmission goes from 59% to nearly 80%. Then it reaches 95% by the end of the simulation. We noticed in Fig. 5 that reward convergence takes more than 800 games in a highly congested network. Experiments with various congestion levels make us realize the need to employ a hybrid communication mode. As illustrated in Fig. 6, when on a highly congested highway, the redundant communication mode is selected more frequently than on a less congested highway. This frequency goes from 3% to 20% with more congestion. Fig. 7 shows a comparison study between four RAT selection strategies. We can see that hybrid communication with a static RAT selection enhances the PRR in both traffic flow densities compared to the baseline performances of ITS-G5 and LTE-V2X.

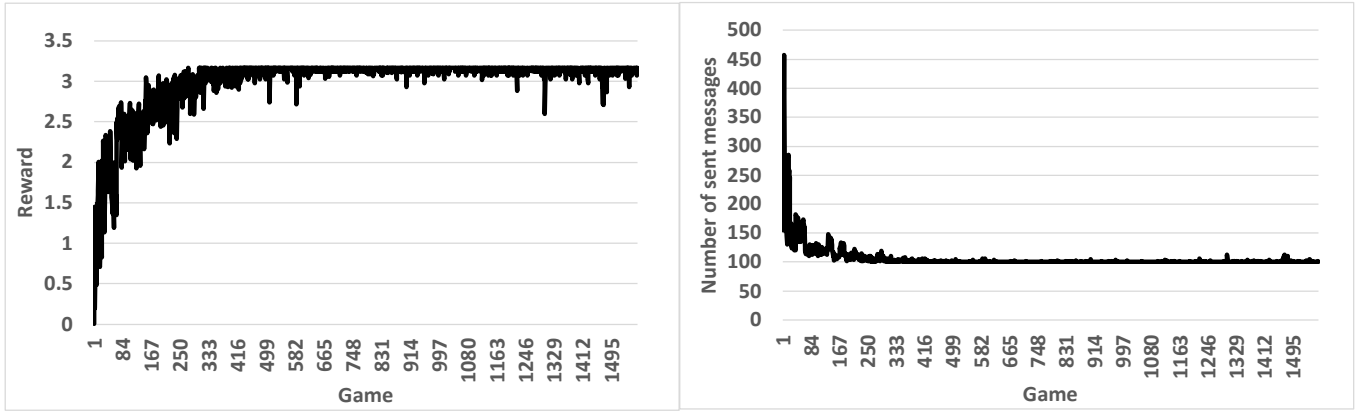

(a) (b)

Fig. 4. The expected reward and the reliability measurements in low congestion level

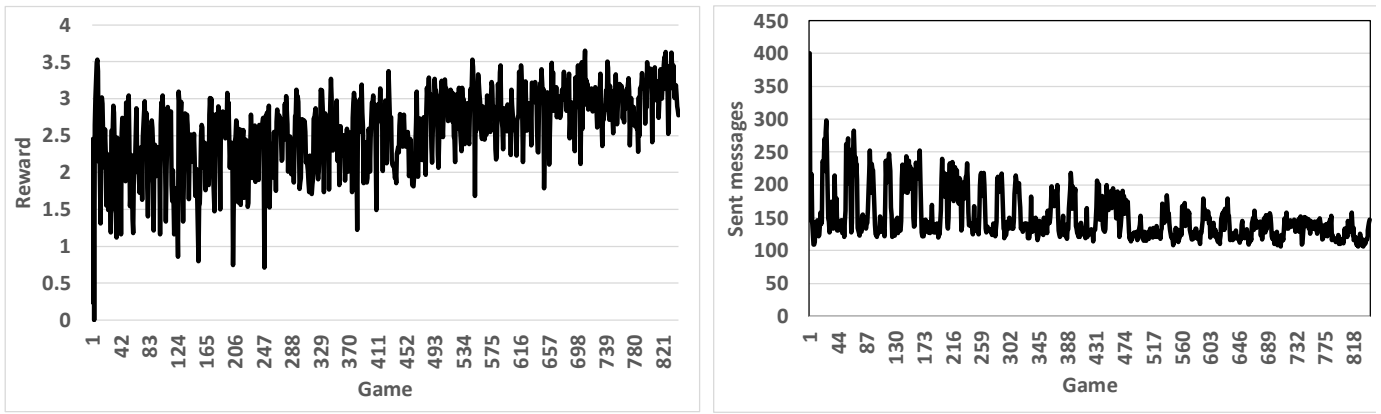

(a) (b)

Fig. 5. The expected reward and the reliability measurements in high congestion level

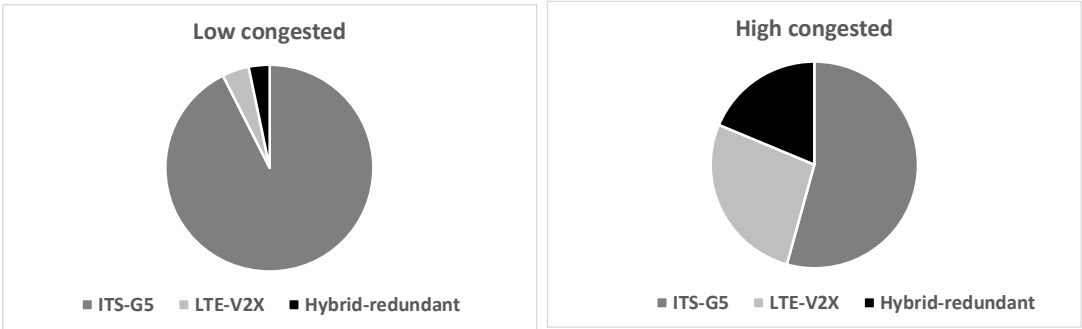

Fig. 6. The percentage use of the redundant communication mode

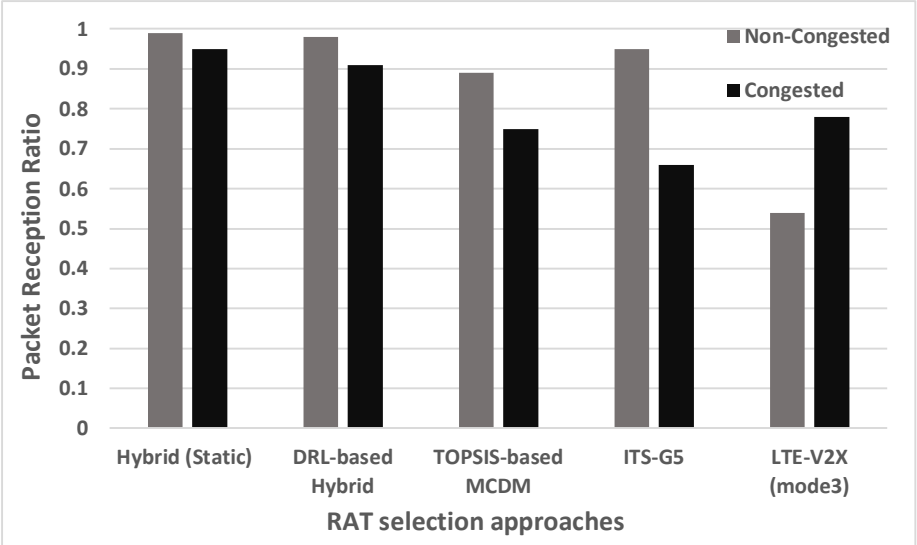

Fig. 7. Comparison between RAT selection approaches

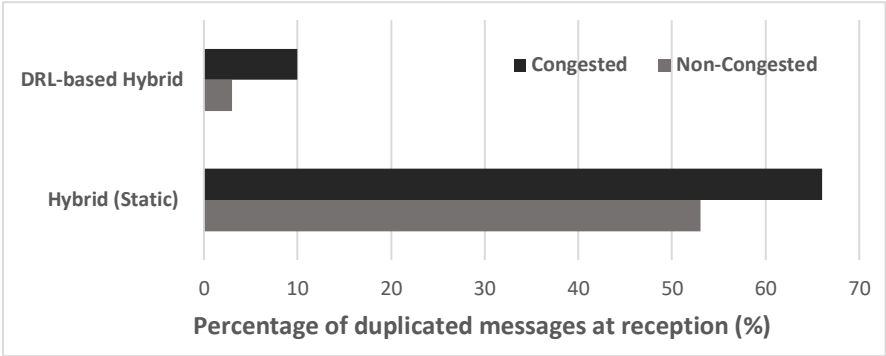

Fig. 8. Percentage of duplicated messages at reception between the two hybrid approaches

The most valuable performance that hybrid communication adds is noticed when the network is congested. We can see that the PRR goes from 66% to more than 90%. We also notice that the TOPSIS-based MCDM RAT selection approach achieves better PRR in low-density (resp. high-density) traffic compared to LTE-V2X (resp. ITS-G5), as explained in [6]. Such results help us notice that these complementary technologies achieve better performance when used together. However, in some cases, we do not need to use the redundant communication mode to meet the requirements of a V2X application. While using static hybrid RAT selection, we report that up to 7119 messages were received two times out of 10718 (during 250s of simulation time), which means 66 % of redundant transmissions were inefficient and considered resource waste. With the help of reinforcement learning, we consider vehicle mobility, the RAT channel quality variations over time, and the application's requirements. As depicted in Fig. 7, our DRL-based communication mode selection method achieves comparable performance to the static hybrid RAT selection approach while minimizing resource waste. According to our duplicated messages statistics in Fig. 8, the DRL-based communication mode selection algorithm decreases the rate of duplicated messages from 66% to nearly 7%. This shows that our selection strategy performs optimally and satisfies the application's requirements without wasting network resources.

## V. Conclusion

This paper implemented and assessed a hybrid vehicular communication architecture with a DRL-based communication mode selection strategy to satisfy recent Day 3 applications' requirements. Performance findings demonstrate that the DRL-based communication mode selection technique improves reliability near 98% while reducing the proportion of duplicate messages to preserve radio resources. We did our best to mimic the network channel access operations of ITS-G5 and LTE-V2X. In future work, we will assess the performance of this selection algorithm using real equipment and explore the coexistence between ITS-G5 and C-V2X in the 5.9 GHz band.